\documentclass[12pt]{article}

\usepackage{epsfig}
\usepackage[inkscapelatex=false]{svg}
\usepackage{physics}
\usepackage{geometry}
\usepackage{graphicx}
\usepackage{subfig}
\usepackage{multirow}
\usepackage{multicol}
\usepackage{caption}
\usepackage{float}
\usepackage{amsfonts}
\usepackage{amsmath}
\usepackage[%
    colorlinks=true,
    pdfborder={0 0 0},
    linkcolor=red]{hyperref}
\usepackage[english]{babel}
\usepackage{cite}
\usepackage{subcaption} 
\usepackage[export]{adjustbox} 
\usepackage{cleveref} 

\usepackage{soul}

\begin{document}
\title{Ultrafast Dynamics of Coherent Phonon Modes in Excitonic Insulator Ta$_2$NiSe$_5$}
\author{Vikas Arora$^{1,2}$, Sukanya Pal$^{1}$, Luminita Harnagea$^{3}$, D. V. S. Muthu$^{1}$, A K Sood$^{*1,2}$}
\date{\today}
\maketitle

\begin{center}
\textit{$^{1}$Department of Physics, Indian Institute of Science, Bangalore 560012, India\\
$^{2}$Center for Ultrafast Laser Applications, Indian Institute of Science, Bangalore 560012, India\\
$^{3}$Department of Physics, Indian Institute of Science Education and Research, Pune, Maharashtra 411008, India \\
}
\end{center}

\begin{center}
\textbf{Keywords:} Excitonic Insulator, Wavelet Transform, Coherent Phonons, optical pump-optical probe spectroscopy, ultrafast carrier relaxation dynamics. 
\end{center}

\begin{abstract}
The spontaneous condensation of excitons in the excitonic insulating phase has been reported in Ta$_2$NiSe$_5$ below $325$ K. In this context, we present the temperature-dependent optical pump optical probe spectroscopy of Ta$_2$NiSe$_5$, with a focus on coherent phonon dynamics. In addition to the fast relaxation process involving excitonic recombination, we observe a systematic behavior for the slow relaxation process associated with the relaxation of hot phonons. The asymmetry parameter and cubic anharmonicity of the 3 THz mode demonstrate the structural transition across $T_C$=325 K, whereas the order parameter nature and asymmetry of 2 THz modes reveal its coupling with the excitonic phase of Ta$_2$NiSe$_5$. Coherent phonon modes display less anharmonicity compared to the corresponding Raman modes. Continuous Wavelet Transform (CWT) reveals that the peak time $t_{peak}$ of phonons is similar for all modes except the 3 THz mode. The temperature dependence of $t_{peak}$ for the M3 mode exhibits a possible role of excitonic condensate below T$_c$ in the formation of quasiparticle (phonon). CWT analysis supports the time-dependent asymmetry of the M3 mode caused by photoexcited carriers. This study illustrates the role of photoexcited carriers in depicting a structural transition and dressing of coherent phonons and, hence, demonstrating many-body effects.
\end{abstract}


\section{Introduction}
An excitonic insulator is an exotic phase of material where electron and hole pairs bind together and condense collectively in the ground state, driven by the relatively weakly screened but attractive Coulomb interaction between valence band holes and conduction band electrons\cite{Halperin1968,Mott1961}. The distinct phase of the material is contingent on two conditions that facilitate the spontaneous creation of excitons: (a) maintaining a sufficiently low electron and hole pair density to allow effective Coulomb interactions while ensuring it is high enough to induce condensation and (b) preventing the rapid recombination of electrons and holes by spatially separating their wavefunctions\cite{Kim2021}. This phenomenon leads to the opening of a bandgap at the Fermi level, reflecting the order parameter of the condensed phase\cite{Pekkar_Varma_2015,Bretscher2021}. In transition to an excitonic insulator (EI), a semiconductor undergoes a Bose-Einstein condensation (BEC), while a semimetal demonstrates a Bardeen-Cooper-Schrieffer (BCS) like transition\cite{Bronold2006,Jerome1967}. In recent years, experimental evidence has emerged in identifying the EI phase, notably, in TmSe$_{0.45}$Te$_{0.55}$\cite{Bucher1991,Wachter2004}, 1\textit{T}-TiSe$_2$\cite{Cercellier2007}, Ta$_2$NiSe$_5$\cite{Wakisaka2009}, twisted heterostructures of monolayer WS$_2$ and bilayer WSe$_2$\cite{Chen2022}, and monolayer WTe$_2$\cite{Jia2022}. The stability of the EI phase in Ta$_2$NiSe$_5$ (TNSe) at room temperature makes it an important material for further studies.
\par
TNSe displays a layered structure, with tantalum and nickel arranged in one-dimensional chains within each layer\cite{Sunshine1985,Kaneko2012}. The spatial separation between Ta(5\textit{d}) conduction band electronic states and Ni(3\textit{d})-Se(4\textit{p}) valence band hole states results in non-overlapping electron and hole wavefunctions. It facilitates the formation of stable excitons, which subsequently undergo condensation. The flatness of the valence and conduction bands resulting from excitonic condensation has been confirmed by angle-resolved photoemission spectroscopy below T=328 K\cite{Wakisaka2009,Wakisaka2012}. The electrical resistivity and X-ray diffraction measurements have shown orthorhombic semiconductor phase to monoclinic excitonic insulator phase on lowering the temperature at 328 K\cite{DiSalvo1986,Kaneko2013}. In Raman spectra, the emergence of the EI state is marked by the appearance of two new phonon modes at 147 cm$^{-1}$ and 235 cm$^{-1}$, along with the splitting of the 119 cm$^{-1}$ mode\cite{Yan2019}. Recent theoretical and experimental studies provide support for the EI transition being of a structural nature, confirmed by spontaneous symmetry breaking arising from phonon instabilities\cite{Baldini2023,Windgätter2021}. However, to support the electronic origin of the transition, it is noted that moving from bulk TNSe to ultrathin (5-layered) TNSe results in no significant changes in the lattice or electronic structure. Nonetheless, the transition temperature is tuned from 325 K to $\sim$300 K, emphasizing the dominance of interband Coulomb interactions over electron-phonon interactions\cite{Kim_2016}. In another report, the presence of all Raman phonon modes across the transition temperature but the emergence of a broad and asymmetric quasi-elastic peak at $\sim$10 cm$^{-1}$ in the vicinity of T$_C$ indicates the electronic origin of EI transition\cite{Kim2021}. Using ultrafast spectroscopy, Werdehausen \textit{et al.} \cite{Werdehausen2018} asserted that TNSe lies within the BEC-BCS crossover regime. This claim was supported by the temperature-dependent behavior of the amplitude and relaxation time of exciton recombination near the transition temperature. However, a temperature-independent gap for all temperatures below the transition was incorporated in the Rothwarf-Taylor model \cite{Rothwarf1967, Kabanov2005} to explain the recombination dynamics, placing TNSe in the BEC regime. In another study, evidence of a coherent amplitude mode (1 THz) in TNSe is presented, analogous to the Higgs mode observed in superconductors\cite{Werdehausen2018r}.
\par
A particle introduced into a many-body system is termed a quasiparticle when it undergoes renormalization by a self-energy cloud, which accounts for the collective effects of many-body interactions\cite{Hase2003}. The coherent oscillations observed in the differential reflectivity in time-resolved pump-probe measurements due to coherent phonon modes can provide insight into the birth of quasiparticles, particularly phonons in this context\cite{Hase2003}. The continuous wavelet transform (CWT) of the oscillatory component of the differential reflectivity signal provides the birth time of a quasiparticle. The intriguing physics of the excitonic insulator TNSe inspired us to investigate the dynamics of both carriers and coherent phonons on ultrafast time scales. Here, we present our findings from optical pump-probe spectroscopy of the excitonic insulator TNSe, performed as a function of temperature. The fast relaxation process is accurately described by the Rothwarf-Taylor model\cite{Rothwarf1967}, incorporating the temperature-independent bandgap. The slower relaxation process exhibits a systematic trend with temperature, indicating the thermalization of phonons. The analysis of coherent oscillations in the differential reflectivity revealed the anomalous behavior of the asymmetric Fano shape of the 3 THz coherent phonon mode across T$_C$= 325K, as well as the order parameter behavior of the 2 THz phonon mode. The CWT spectrogram reveals coincident birth times for all phonon modes except the 3 THz mode. The temperature dependence of the birth time for 3 THz mode demonstrates the role of excitonic condensate of TNSe across $T_C$. In the end, CWT analysis reveals that the origin of asymmetry in 3 THz mode is temporal, specifically attributed to photoexcited carriers in the background. To the best of our knowledge, apart from a study on pyrochlore titanates \cite{Kamaraju2011}, there are no reports comparing coherent phonons in pump-probe experiments with the corresponding phonons observed in Raman spectroscopy. Interestingly, we observe that coherent phonon modes in TNSe exhibit less anharmonicity compared to Raman phonon modes.


\section{Experimental details}
Ta$_2$NiSe$_5$ (TNSe) single crystals were grown by the chemical vapor transport method, utilizing iodine as the transport medium, as described in \cite{SukanyaPal2020}. Temperature-dependent optical pump-probe spectroscopy was performed using liquid nitrogen cooled Linkam THMS350V stage across a temperature range from 80 K to 380 K, with temperature accuracy of $\sim$0.1 K. Pump-probe experiments were done using $\sim$50 fs, 1 kHz amplifier (Spectra-Physics, Spitfire Ace) with a central wavelength of 800 nm. The fluence of the pump beam was kept fixed at 556 $\mu J/cm^2$ across varying temperatures. Raman spectra were recorded using a LabRam spectrometer (M/s Horiba) in back-scattering geometry, using 50X objective and laser excitation of 532 nm from a diode-pumped solid-state laser, with laser power of $\sim$3 mW.

\section{Results and discussions}
Figure \ref{Fig:5p1}\textcolor{red}{a} shows the time-resolved differential reflectivity ($\Delta$R/R) up to 10 ps delay at a temperature of T=180 K  while the inset shows the data till delay time of 120 ps. The coherent oscillations observed in the $\Delta$R/R represent the coherent phonon modes, which are extracted by subtracting the electronic relaxation comprising of both fast and slow decay processes, along with a long-lasting component, fitted using
\begin{equation}
    \centering
    \frac{\Delta R}{R}(t)=\frac{1}{2}(1+erf(\frac{t-t_0}{\tau_{rise}}))(A_1e^{-t/\tau_1}+A_2e^{-t/\tau_2}+A_3)
    \label{eq:5p1}
\end{equation}
where $A_1$ ($A_2$) and $\tau_1$ ($\tau_2$) represent the amplitude and relaxation time for the fast (slow) dynamical process, respectively. A$_3$ denotes the long-lived bolometric signal, demonstrating the dissipation of heat from the irradiated area, which decays over $\sim$10 ns delay time, thus maintaining a constant value on a 100 ps time scale. Our results are qualitatively similar to the published reports\cite{Demsar2006,Werdehausen2018,Werdehausen2018r}. The validity of the model is demonstrated in Figure \ref{Fig:5p1}\textcolor{red}{a}, where the experimental data for $\Delta$R/R (represented by a blue curve) is well-fitted with the biexponential decay (shown as a solid red curve). The inset illustrates the good fit of the data up to 120 ps. The bottom section of Figure \ref{Fig:5p1}\textcolor{red}{a} displays the coherent oscillations obtained by subtracting the biexponential fitting model (Eq.\ref{eq:5p1}, the red solid curve) from the time-resolved $\Delta$R/R signal (the blue curve). Before we delve into the understanding of these oscillations, we would first discuss the electronic relaxation dynamics with temperature. 

\subsection{Electronic Relaxation}
\begin{figure}[H]
   \centering
   \includegraphics[width=\textwidth]{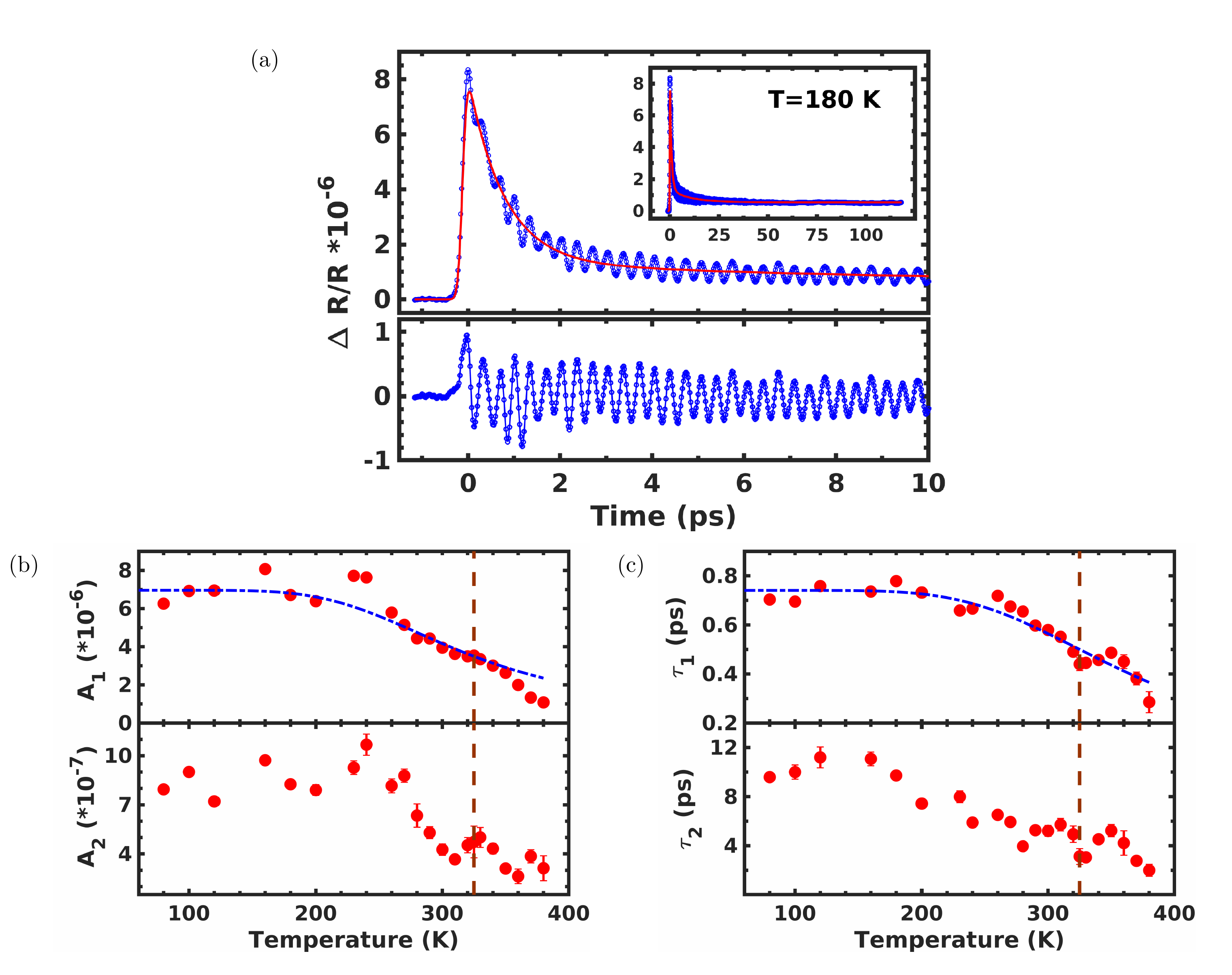}
   \caption[Differential reflectivity of Ta$_2$NiSe$_5$ and relaxation dynamics' parameters as a function of temperature]{(a) The differential reflectivity ($\Delta R/R$) for TNSe is presented with delay time (solid blue circles) at T=180 K and fluence of 556 $\mu$J/cm$^2$, the red solid curve depicts the biexponential model (Eq.\ref{eq:5p1}). The inset demonstrates the validity of the biexponential model up to 120 ps. The graph at the bottom represents the coherent oscillation in the real-time $\Delta$R/R. (b) The amplitude, $A_1$ and $A_2$ and (c) the corresponding relaxation times, $\tau_1$ and $\tau_2$ of the fast and slow relaxation processes described by Eq.\ref{eq:5p1}, are presented at different temperatures. The fast relaxation process, characterized by $A_1$ and $\tau_1$ is effectively demonstrated by the RT model (dashed blue curves) up to the transition temperature, $T_C$=325 K, with $\Delta$=131 meV. The brown dashed line shows the transitional behavior across $T_C$.}
    \label{Fig:5p1}
\end{figure}

In a previous report on ultrafast spectroscopy of TNSe \cite{Werdehausen2018}, the amplitude A$_1$ and the relaxation time $\tau_1$ remain constant up to $\sim$200 K, after which they decrease with increasing temperature. The amplitude A$_2$ and $\tau_2$ exhibited a constant behavior with temperature, except for a jump in the constant value of $\tau_2$ above $\sim$200 K. Figure \ref{Fig:5p1}\textcolor{red}{b} and \ref{Fig:5p1}\textcolor{red}{c} displays the fitting parameters ($A_1$,$A_2$) and ($\tau_1$,$\tau_2$) at different temperatures. The amplitude ($A_1$) and the corresponding relaxation time ($\tau_1$) of the fast relaxation process, represented by solid red circles, exhibit a constant value till $\sim$250 K, followed by a gradual fall with further increase in temperature, similar to \cite{Werdehausen2018}. However, the temperature dependence of A$_2$ and $\tau_2$ shown in Figure \ref{Fig:5p1}\textcolor{red}{b} and \ref{Fig:5p1}\textcolor{red}{c} is different from earlier report \cite{Werdehausen2018}. The amplitude $A_2$ shows a similar temperature dependence as that of A$_1$. The corresponding relaxation time $\tau_2$ shows a systematic fall with increase in temperature. Werdehausen \textit{et al.} \cite{Werdehausen2018} considered the number of photoexcited carriers above the excitonic gap to be directly related to amplitude A$_1$(T), with their relaxation governed by $\tau_1$(T). Both parameters are derived using the Rothwarf-Taylor (RT) model with a temperature-independent excitonic gap \cite{Rothwarf1967}. Physically, the RT model incorporates the disintegrating of the excitons into unbounded electrons and holes by the pump pulse. Following the photoexcitation, the excited electrons and holes recombine to form excitons, resulting in the emission of high-frequency phonons (HFPs). The recombination described here corresponds to the fast relaxation process ($A_1$, $\tau_1$). There are now two possibilities: (a) The HFPs released in this process can either disintegrate another exciton to produce an electron and a hole, or (b) decay into low-frequency phonons (LFP), which are incapable of further excitonic dissociation. As the excitonic insulator to a semiconducting phase transition temperature is approached, the process (a) can instigate an avalanche of free electrons and holes, leading to a bottleneck for them to recombine into an exciton. This phenomenon results in a substantial increase in relaxation time near the transition point\cite{Kabanov1999,Kabanov2005}. For the process (b) the decay of a high-frequency phonon (HFP) into low-frequency phonons occurs much more rapidly than the creation of a fermion pair\cite{Torchinsky2010}. The observed decrease in the fast relaxation time ($\tau_1$) as the transition temperature $T_C$ is approached (Figure \ref{Fig:5p1}\textcolor{red}{c}), illustrates the process (b) where the HFP decays into LFPs. In addition to qualitative understanding, acquiring a quantitative estimation of the amplitude and relaxation time with temperature would provide more comprehensive insights.
\par
Utilizing the RT model, the expressions for the amplitude ($A_1$) and the corresponding relaxation time ($\tau_1$) can be derived, taking into account the temperature-independent gap (excitonic binding energy, $\Delta_{EI}\equiv2\Delta$)\cite{Kabanov1999,Kabanov2005,Werdehausen2018}.
\begin{align}
    A_1(T)=\frac{A_0/\Delta}{1+Jexp(-\frac{\Delta}{k_BT})} &&   \tau_1(T)=\frac{1}{K+L\sqrt{\Delta T}exp(-\frac{\Delta}{k_BT})}
    \label{eq:5p2}
\end{align}
where A$_0$, J, K, L and $\Delta$ are fitting parameters. The blue dashed curve in Figures \ref{Fig:5p1}\textcolor{red}{b} and \ref{Fig:5p1}\textcolor{red}{c} accurately describes the behavior of $A_1$ and $\tau_1$ respectively till T$_C$=325 K. Our fitting process gives $\Delta = 131$ meV, corresponding to $\Delta_{EI}$ = 262 meV, which aligns closely with the estimated range of 250 to 320 meV\cite{Tang2020,Wakisaka2012,Lu2017}. Thus, the Rothwarf-Taylor model effectively describes the condensed phase of excitons in TNSe. When comparing the amplitudes $A_1$ and $A_2$ (Figure \ref{Fig:5p1}\textcolor{red}{b}) for T$<$$T_C$, $A_1$ is one order of magnitude larger than $A_2$, suggesting that the carrier dynamics is primarily governed by the quasiparticles recombining into excitons. Consistent with ref. \cite{Werdehausen2018}, we attribute the slow component in Eq.\ref{eq:5p1}, namely $A_2$ and $\tau_2$, to thermalization of phonons (HFPs). The relaxation time $\tau_2$ signifies the process of hot HFPs gradually cooling down by decaying into low-frequency phonons (LFPs). As temperature increases, the expanded phase space available for phonons results in a more rapid cooling process, leading to a decrease in $\tau_2$ with temperature. In the semiconductor region, for temperatures T$>$$T_C$, the relaxation dynamics are characterized by photoexcited carriers transferring energy to optical phonons, which subsequently relax through low-energy acoustic phonons. The fast relaxation time ($\tau_1$) pertains to the cooling of carriers by emission of optical phonons ($<$1 ps), while the slow relaxation time ($\tau_2$) relates to further cooling of optical phonons via acoustic phonons ($\sim$10 ps). 

\subsection{Coherent Phonons}
\begin{figure}[H]
  \centering
  \includegraphics[width=\textwidth]{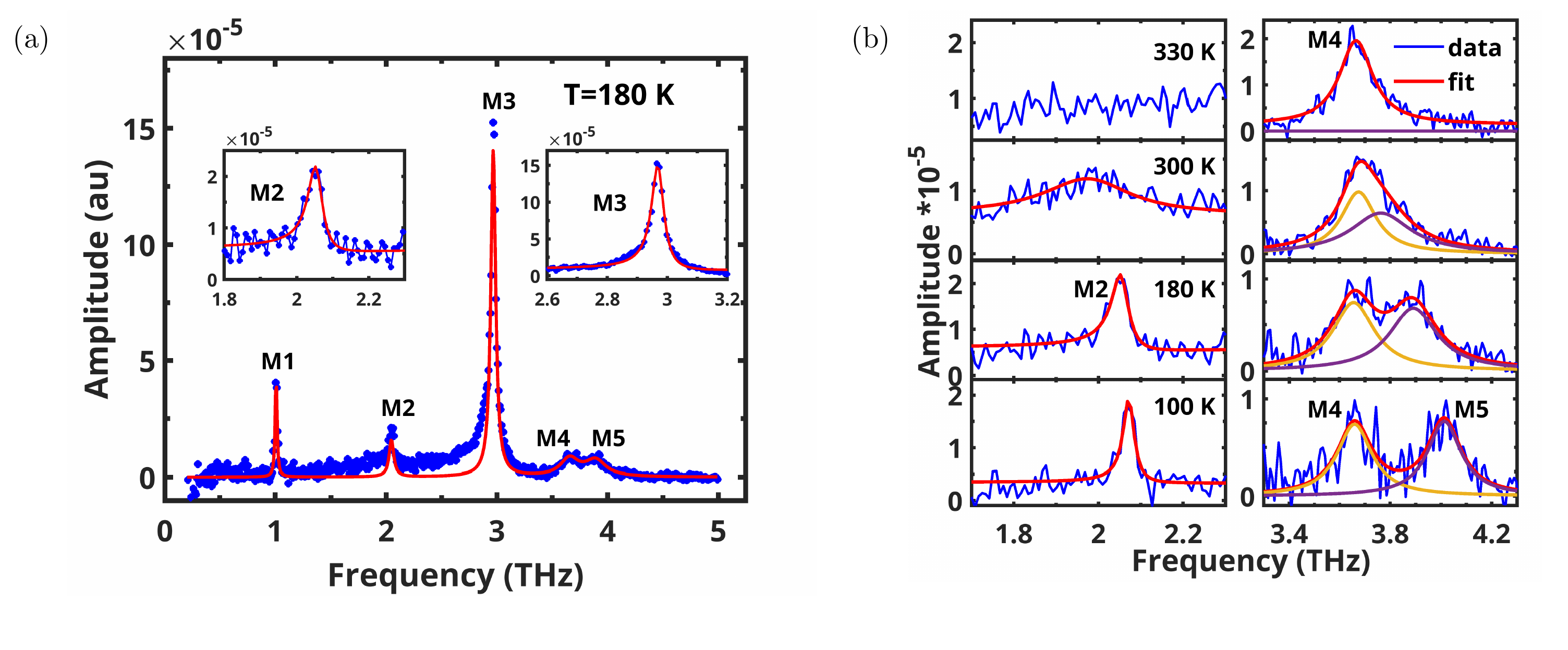}
  \caption[Coherent phonon mode analysis for EI Ta$_2$NiSe$_5$ as a function of temperature]{(a) The coherent phonon modes of TNSe, represented by solid blue circles, are fitted by Lorentzians, depicted as a red solid curve. The modes are labeled as follows: M1=1 THz, M2=2 THz, M3=3 THz, M4=3.65 THz, and M5=4 THz. M2 and M3 exhibit disparity with the Lorentzian fit. Insets depict FANO fits for the M2 and M3 modes. (b) The disappearance of M2 (left panel) and merging of M4 and M5 (right panel) modes across T$_C$.} 
\label{Fig:5p2}
\end{figure}

\begin{figure}[H]
    \centering
    \includegraphics[width=\textwidth]{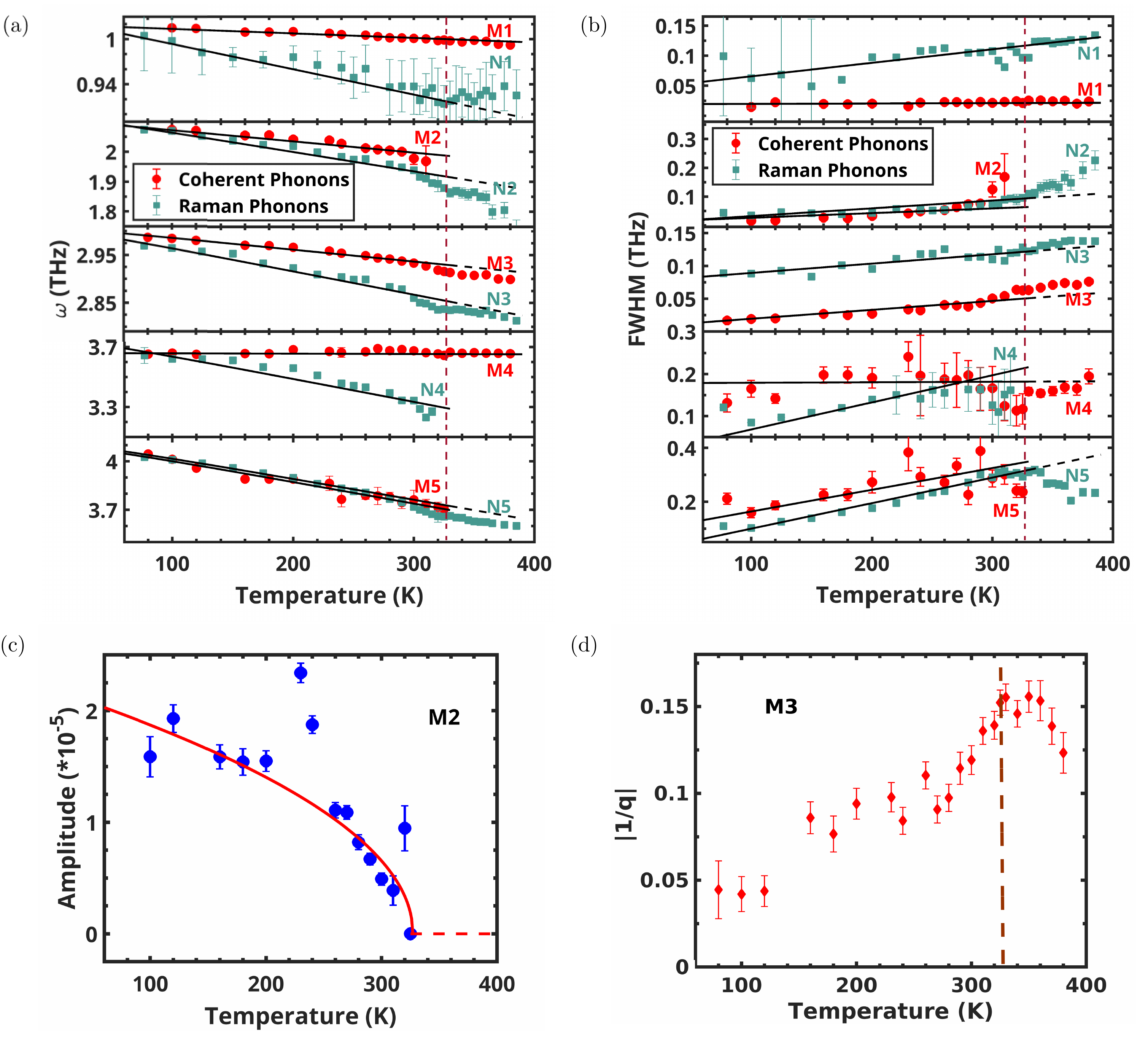}
    \caption[Amplitude of coherent phonon mode M2 and asymmetry parameter of coherent phonon mode M3 with temperature]{(a) The peak frequency, $\omega$, and (b) Full Width at Half Maximum (FWHM) of coherent phonons (red circles) and Raman phonons (teal squares) are plotted at different temperatures for all the modes. (c) The amplitude of the coherent phonon mode M2 displays order parameter behavior, vanishing beyond $T_C$. (d) The temperature variation of the asymmetry parameter $|1/q|$ for the M3 mode is presented across the transition temperature, $T_C$. It exhibits a systematic increase as the temperature rises until reaching the transition point.}
    \label{Fig:5p3}
\end{figure}

The coherent oscillations observed in the lower panel of Figure \ref{Fig:5p1}\textcolor{red}{a} allow us to extract coherent phonon modes using the Fast Fourier Transform (FFT). Figure \ref{Fig:5p2}\textcolor{red}{a} represents the coherent phonon modes, including M1 (1 THZ), M2 (2THz), M3 (3THz), M4 (3.65 THz), and M5 (4.0 THz) depicted by the solid blue circles, is consistent with findings from earlier Raman studies\cite{Mor2018,Yan2019}. The phonon modes have been fitted using a sum of Lorentz functions, represented by red solid curve. As can be seen in Figure \ref{Fig:5p2}\textcolor{red}{a}, modes M1, M4, and M5 were fitted well by Lorentzian functions, while M2 and M3 modes exhibit asymmetric lineshapes and were consequently fitted to the Breit-Wigner-Fano (BWF) lineshape, as illustrated in the insets. The BWF lineshape arising from quantum interference between discrete and continuum transitions is given by $I(\omega)=I_0\frac{[1+(\omega-\omega_{F})/q\Gamma]^2}{1+[(\omega-\omega_{F})/\Gamma]^2}$
where $|1/q|$ represents an asymmetry factor characterizing the coupling strength between the discrete (phonons in this case) and continuous transitions (continuum of photoexcited carriers)\cite{Fano2016,miroshnichenko2010}. The parameter $\Gamma$ is the broadening parameter, while $\omega_{F}$ corresponds to the uncoupled BWF peak frequency. At the transition temperature $T_C$, mode M2 vanishes, while modes M4 and M5 coalesce, as depicted in Figure \ref{Fig:5p2}\textcolor{red}{b}. Figure \ref{Fig:5p3}\textcolor{red}{a} and \ref{Fig:5p3}\textcolor{red}{b} show the peak frequency and full width at half maximum (FWHM) for various modes as a function of temperature. The solid black lines represent the simplistic cubic anharmonicity model for the peak frequency, expressed mathematically as 
\begin{equation}
    \omega_{cubic}(T) = \omega_0 - A[1 + 2n(\omega_0/2)]
    \label{eq:5p3}
\end{equation}
and for the full width at half maximum (FWHM), presented as
\begin{equation}
    \Gamma_{cubic}(T) = \Gamma_0 + B[1 + 2n(\omega_0/2)]
    \label{eq:5p4}
\end{equation}
In this simple anharmonic model, the phonon with frequency $\omega_0$ decays into two phonons of equal frequency ($\omega_0/2$) \cite{Klemens_1966}. Note that the parameters A and B are both positive. The temperature dependence of mode M3 for $T > T_C$ deviates from the extrapolated values of the above model, as is expected for a structural transition.
\par
In the BEC condensate phase of TNSe, a mean-field-like order parameter behavior has been reported for the signal amplitude, represented by the maximum amplitude of $\Delta$R/R \cite{Bretscher2021}. Werdehausen \textit{et al.} \cite{Werdehausen2018r} and Mor \textit{et al.} \cite{Mor2018} presented the amplitude of the coherent phonon modes M1 and M4 exhibiting order parameter behavior. In comparison to these phonon modes, we observe the order parameter for the amplitude of the coherent phonon mode M2, as seen by the red line depicting Amplitude $\sim$ A$_{11}\sqrt{T-T_C}$ (Figure \ref{Fig:5p3}\textcolor{red}{c}). Figure \ref{Fig:5p3}\textcolor{red}{d} shows the temperature dependence of $|1/q|$ for the M3 mode, displaying that the coupling strength is maximum at $T_C$. The coupling strength $|1/q|$ for the M2 mode decreases as T$_C$ is approached, as shown in Figure \ref{Fig:5p6}\textcolor{red}{b}. The absence of the M2 mode at $T>T_C$, along with its order parameter-like trend, suggests that it is coupled to the condensed excitonic phase of TNSe. 

\subsection{Time-evolution of phonon modes}
\begin{figure}[H]
    \centering
    \includegraphics[width=\textwidth]{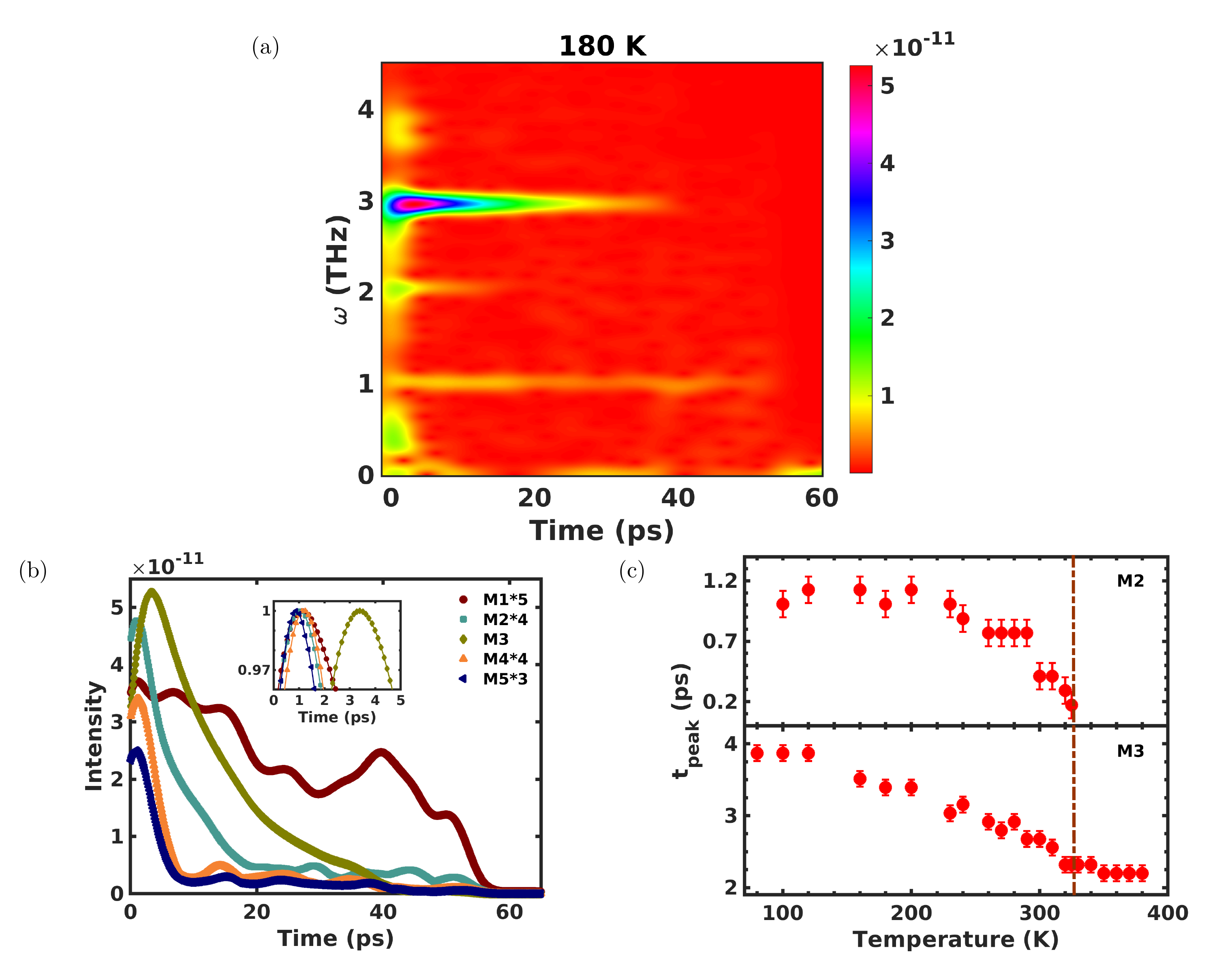}
    \caption[Continuous wavelet (CWT) analysis of coherent phonon modes]{The Continuous Wavelet Transform (CWT) Analysis (a) CWT spectrogram of coherent oscillations (Figure \ref{Fig:5p1}\textcolor{red}{a}) represents the evolution of different coherent phonon modes with time at 180 K. (b) Time evolution of the intensity of all modes at 180 K, with the inset displaying that the peak time ($t_{peak}$) of M3 exceeds that of all other modes. The intensity of the modes is scaled for better comparison. (c) The peak time ($t_{peak}$) of M2 and M3 is plotted against temperature.}
    \label{Fig:5p4}
\end{figure}

While coherent phonon modes offer a plethora of information about TNSe, studying their time evolution can offer insights into the dressing of phonons due to photoexcited carriers, known as birth time \cite{Hase2003}, and the time evolution of asymmetry of phonon modes. 
The continuous wavelet transform (CWT) is a tool to describe the temporal evolution of coherent phonons. The oscillatory component of the differential reflectivity (depicted in the bottom section of Figure \ref{Fig:5p1}\textcolor{red}{a}) is analyzed using continuous wavelet transform (CWT), following \cite{Hase2003}, utilizing the Morlet mother wavelet as follows \cite{Kamaraju_2010}:
\begin{equation}
    \psi(t/s)=\pi^{-1/4}e^{-t^2/2s^2+jt/s} \\ \label{eq:5p5}
\end{equation}
The amplitude of the continuous wavelet transform (CWT) for a fixed scaling factor $s$ at a given time $\tau$ is expressed as:
\begin{equation}
    CWT(\tau,s)=\frac{1}{\sqrt{|s|}}\int\frac{\Delta R}{R}(t)\psi^*(\frac{t-\tau}{s})dt \label{eq:5p6}
\end{equation}
In simpler terms, the CWT coefficient for a fixed scaling factor $s$ is obtained by taking the Fast Fourier Transform (FFT) of the product of the real-time signal and a Gaussian envelope, where the envelope slides in time with a step size of $\tau$. The CWT spectrogram for TNSe at T = 180 K is illustrated in Figure \ref{Fig:5p4}\textcolor{red}{a}, presenting the temporal evolution of modes M1 to M5. The time evolution of all the modes is displayed in Figure \ref{Fig:5p4}\textcolor{red}{b}, with appropriate scaling as indicated in the legends, for better visualization. The inset shows the normalized intensity for the time evolution of modes, highlighting the birth of these quasiparticles, phonons, defined as the time when the intensity of a particular mode reaches its maximum, denoted by $t_{peak}$. It is evident that $t_{peak}$ is approximately the same for all modes except for the strongest mode M3. The observation of higher $t_{peak}$ for the stronger mode is also seen in coherent phonon A$_{1g}^1$ in Bi$_2$Te$_3$ \cite{Kamaraju_2010}. We selected two modes, M2 and M3, with a better signal-to-noise ratio to analyze the evolution of $t_{peak}$ with temperature, as depicted in Figure \ref{Fig:5p4}\textcolor{red}{c}. For temperatures below $260$ K, mode M2 exhibits a constant value for $t_{peak}$, which then gradually decreases as the transition temperature is approached. The $t_{peak}$ for M3 mode shows a gradual fall with an increase in temperature until reaching the transition temperature $T_C$, beyond which it remains constant. With the increase in temperature, the dressing of M3 mode occurs sooner due to the availability of additional phonon phase space. However, the change in slope of $t_{peak}$ for M3 across T$_C$ illustrates that the birth of coherent phonons is affected by the condensate. We want to emphasize that the absolute value of $t_{peak}$ depends on the scaling factor, $s$, and sliding time, $\tau$. We have identified an optimized value for both parameters to facilitate the presentation of CWT analysis. Thus, while the absolute value of $t_{peak}$ may vary, its trend with temperature remained independent of these parameters. Next, we address the asymmetry of the M3 mode through the CWT analysis.
\par
\begin{figure}[H]
    \centering
    \includegraphics[width=\textwidth]{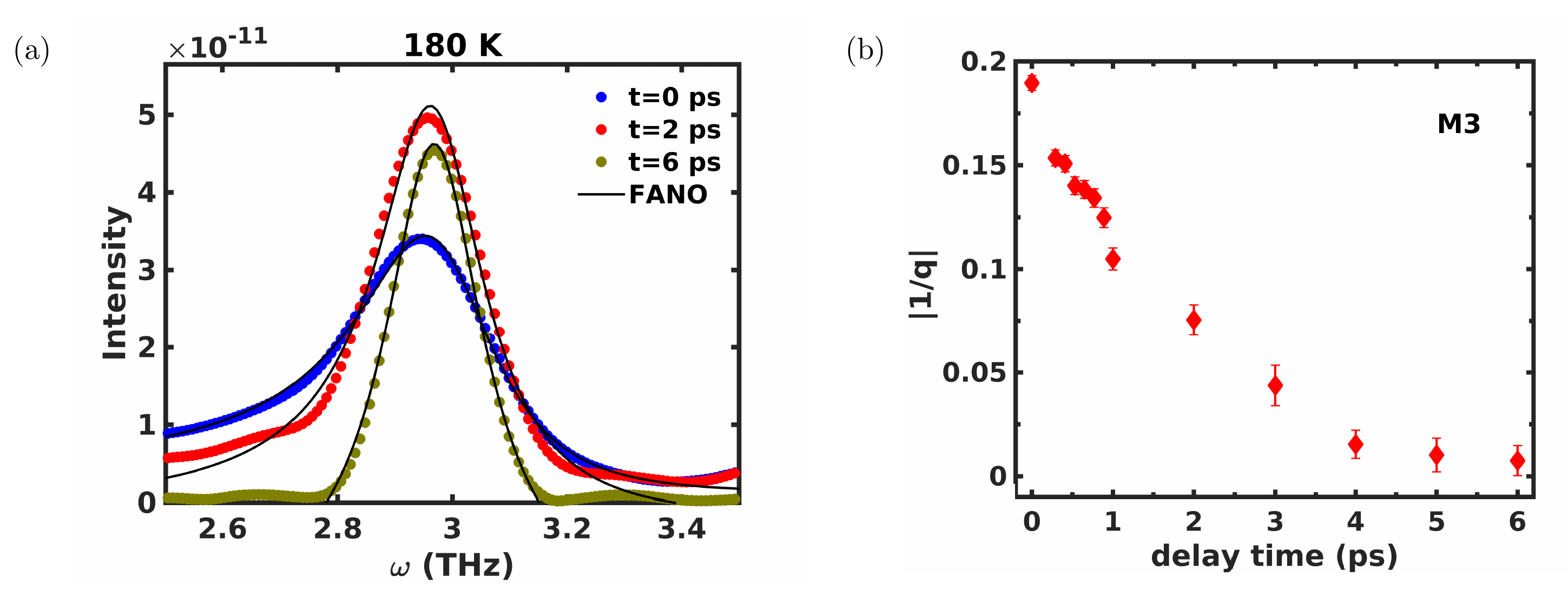}
    \caption[Time evolution of Fano lineshape for coherent phonon mode M3]{(a) The spectrogram snapshots of the M3 mode at t=0 ps, 2 ps, and 6 ps (solid circles) and the black curve (BWF lineshape) demonstrate that the asymmetric nature of the mode decreases with time. (b) The asymmetry parameter is plotted against delay time, showing a decrease with time and approaches zero beyond 4 ps.}
    \label{Fig:5p5}
\end{figure}

The vertical snippets of the CWT spectrogram (Figure \ref{Fig:5p4}\textcolor{red}{a}) at various delay times offer insights into the evolution of phonon modes over time. Figure \ref{Fig:5p5}\textcolor{red}{a} presents the behavior of M3 mode at three distinct time points, t=0 ps, 2 ps, and 6 ps. The asymmetric nature of the M3 mode decreases with time evolution, represented by the black curve. Figure \ref{Fig:5p5}\textcolor{red}{b} shows the asymmetry parameter $|1/q|$ versus delay time. As time evolves, $|1/q|$ decreases and becomes almost negligible beyond t=4 ps. Hence, it is evident that the asymmetry of the M3 mode arises from its coupling with the photoexcited carriers in the background. As these carriers relax to the ground state, forming excitons, the asymmetry of the M3 mode disappears. Given the diverse temperature-dependent behavior of different coherent phonon modes, it would be valuable to investigate the Raman counterparts of these phonon modes, and compare the results.

\subsection{Raman results}

\begin{figure}[H]
    \centering
    \includegraphics[width=\textwidth]{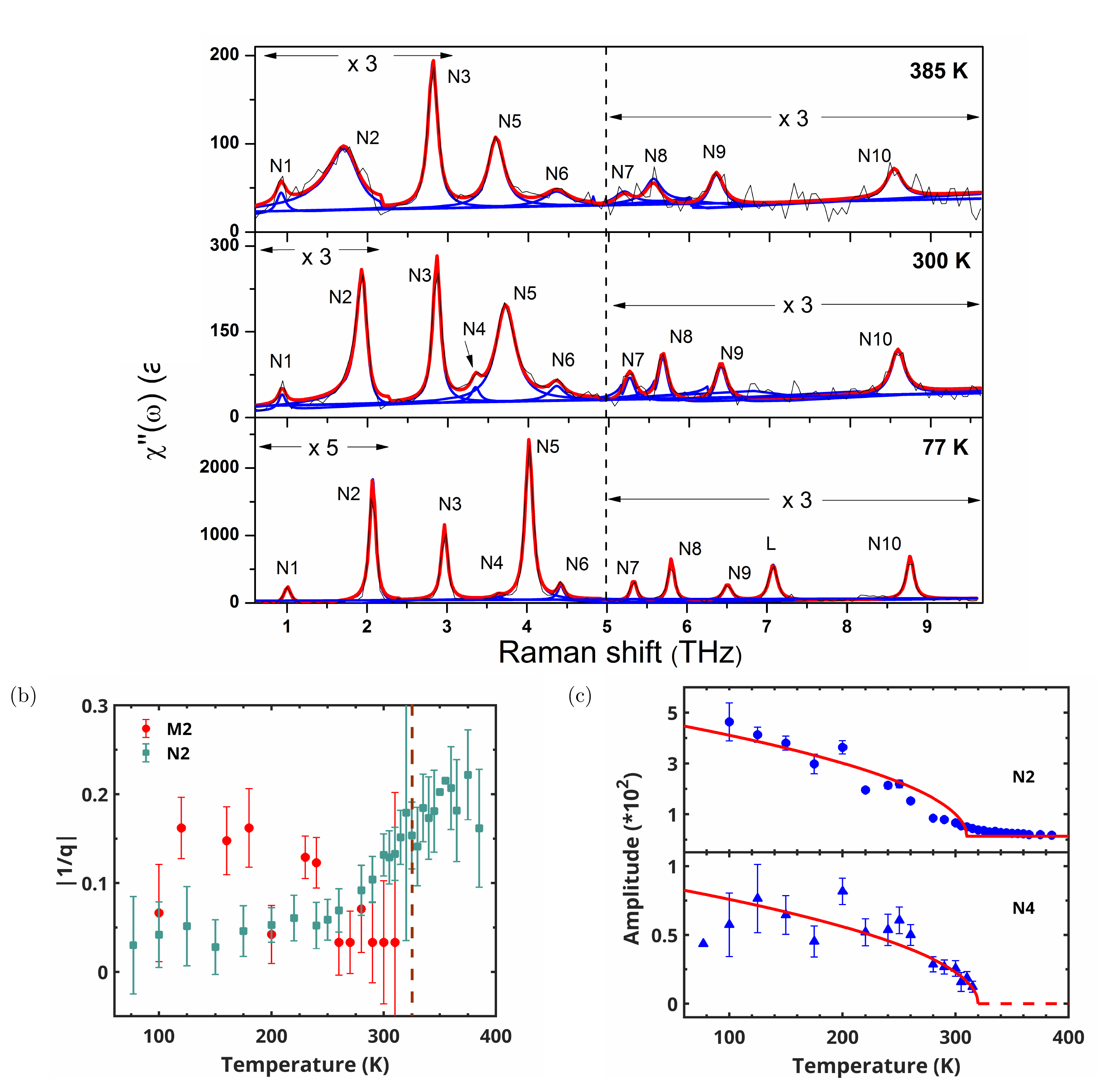}
    \caption[Temperature evolution of Raman spectra of Ta$_{2}$NiSe$_{5}$]{(color online) Temperature evolution of Raman spectra of Ta$_{2}$NiSe$_{5}$. The solid lines (red and blue) are the Lorentzian fits to the experimental data(black). The Raman mode N4 indicated by the arrow disappears above T$_{c}$ in temperature-induced phase transitions. (b) The asymmetry of coherent phonon M2 and Raman phonon N2 is presented with temperature, (c) The amplitude of Raman modes N2 and N4 show order parameter behavior.}
    \label{Fig:5p6}
\end{figure}

There is a total of 45 optical phonons at the Brillouin zone center of Ta$_{2}$NiSe$_{5}$, among which 21 modes are infrared active [11B$_{1u}$+10A$_{u}$] and 24 are Raman active [11A$_{g}$ + 13B$_{1g}$] \cite{Werdehausen2018r}. Figure \ref{Fig:5p6}\textcolor{red}{a} shows the temperature evolution of the reduced Raman susceptibility $\chi$\textquotedblright$(\omega)$ = Intensity($\omega$)/(n($\omega$)+1), of Ta$_{2}$NiSe$_{5}$ at a few typical temperatures where n($\omega$) is the Bose-Einstein population factor. We observe 11 modes at 77 K: N1($A_{g}$ at 0.99 THz), N2($B_{1g}$ at 2.01 THz), N3($B_{1g}$ at 3 THz), N4($B_{1g}$ at 3.66 THz), N5($A_g$ at 3.99 THz), N6($A_{g}$ at 4.41 THz), N7($A_{g}$ at 5.31 THz), N8($B_{1g}$ at 5.78 THz), N9($B_{1g}$ at 6.50 THz) and N10($B_{1g}$ at 8.78 THz). To distinguish between Raman and coherent phonon modes, the Raman phonon modes are denoted with the letter `N', while the coherent phonon modes are expressed as `M'. The symmetry assignments are taken from a recent study (supplementary materials of\cite{Werdehausen2018r}). Mode marked as L ($B_{1g}$ at 7.10 THz), becomes very weak with increasing temperature and cannot be clearly resolved above 200 K. Loretzian functions were fitted to Raman susceptibility $\chi$\textquotedblright$(\omega)$ to extract the phonon frequencies and full width at half maximum (FWHM) of the Raman modes, except for mode M2, which shows asymmetric lineshape and hence was fitted to BWF lineshape.
\par
We will now comment on the reported Raman results. Yan \textit{et al.} \cite{Yan2019} have reported the merging of N4 and N5 and the disappearance of N6 and L modes beyond T$_C$ of 328 K. In comparison, Kim \textit{et al.} \cite{Kim2021} reported that N5, though absent in unpolarized spectra, is present in polarization-resolved spectra. They showed the presence of all the 11 phonon modes across the transition temperature of 325 K, claiming the electronic origin of the EI transition in TNSe. We observed the presence of the N2 mode across T$_C$ (Figure \ref{Fig:5p6}\textcolor{red}{a}), in contrast to the disappearance of the coherent M2 mode above T$_C$. Figure \ref{Fig:5p6}\textcolor{red}{a} shows the absence of Raman phonon mode N4, but the presence of mode N5 above T$_C$, whereas in the time-resolved spectroscopy, coherent phonon mode M5 disappears instead of M4, as supported by previous Raman reports \cite{Yan2019}. Figures \ref{Fig:5p3}\textcolor{red}{a} and \ref{Fig:5p3}\textcolor{red}{b} present the temperature dependence of phonon frequencies ($\omega$) and FWHM ($\Gamma$). The solid black lines are fit to the cubic anharmonic model, Eq. \ref{eq:5p3} and \ref{eq:5p4}. The deviations from anharmonicity for all the phonon modes for temperatures above T$_C$ depict the structural transition for TNSe. Above T$_{c}$, the FWHM of mode N5 shows an anomalous decrease, indicating phonon dynamics influenced by strong electron-phonon coupling (Figure \ref{Fig:5p3}\textcolor{red}{b}). The increase in the broadening parameter ($\Gamma$) (Figure \ref{Fig:5p3}\textcolor{red}{b}) and the Fano asymmetry parameter $|1/q|$ with temperature (Figure \ref{Fig:5p6}\textcolor{red}{b}) reflect the enhanced electron-phonon coupling with temperature. The amplitude of the N2 and N4 modes follows order parameter behavior (Figure \ref{Fig:5p6}\textcolor{red}{c}), consistent with coherent M2 mode.

\subsection{Comparison between Coherent phonons and Raman phonons}

\begin{figure}[H]
    \centering
    \includegraphics[width=\textwidth]{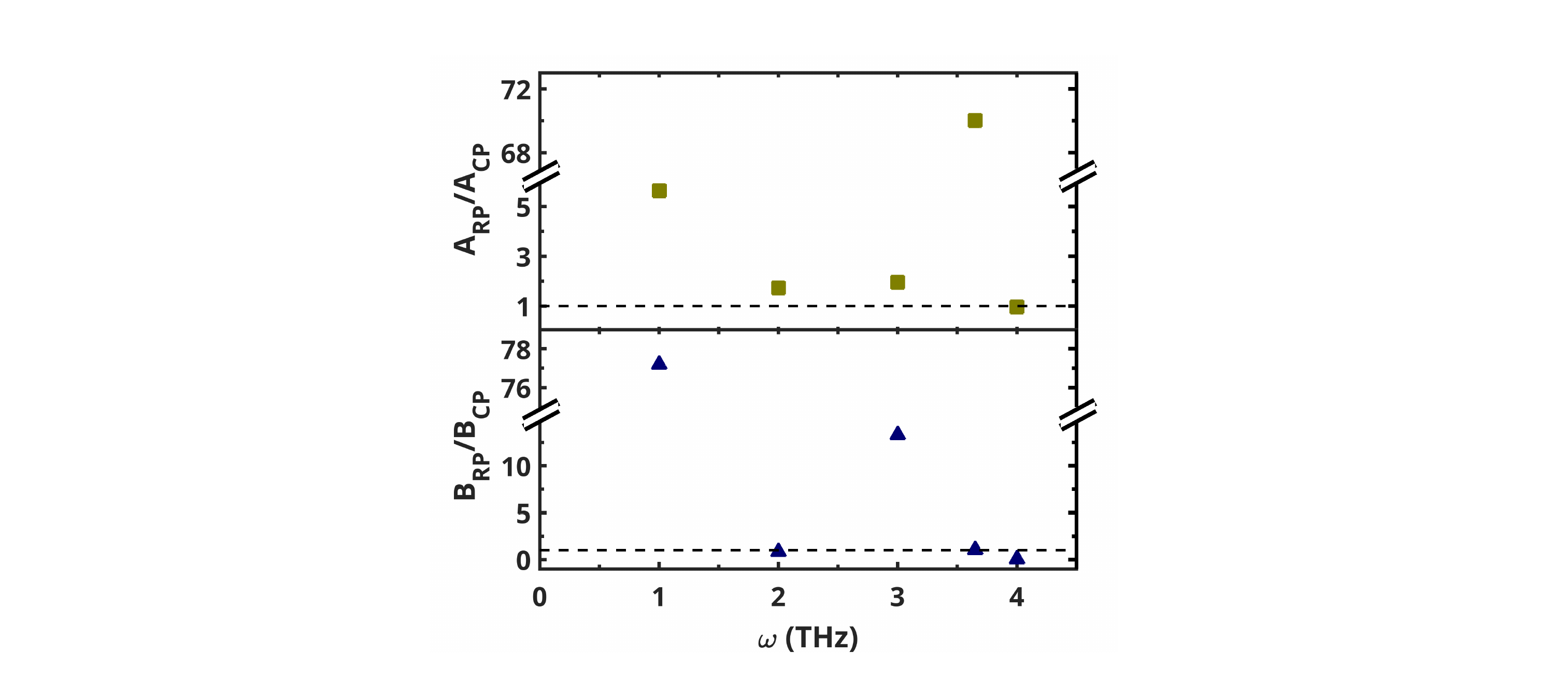}
    \caption[Anharmonicity comparison for Raman and coherent phonon modes of Ta$_2$NiSe$_5$]{The top (bottom) panel shows the ratio of anharmonic coefficient, A (B) (Eq.\ref{eq:5p3}(\ref{eq:5p4})) for Raman and coherent phonons for TNSe.}
    \label{Fig:5p7}
\end{figure}

The origin of coherent phonons in opaque materials is governed by the mechanism of displacive excitation of coherent phonons (DECP). In this, the pump pulse induces a sudden onset of photoexcited carriers, displacing the equilibrium position of lattice atoms, and consequently, these atoms start oscillating around their new mean position\cite{Zeiger1992,Mazin1994}. Figure \ref{Fig:5p3}\textcolor{red}{a} shows that the mode frequency ($\omega$) for Raman phonons (RP) is less than the coherent phonons (CP) across the temperature range of 80 K to 380 K. 
Figure \ref{Fig:5p7} (top panel) shows that the ratio of the anharmonicity coefficient (parameter A in Eq.\ref{eq:5p3}) of Raman phonons, A$_{RP}$ to the coherent phonons, A$_{CP}$ is greater than one for all the phonon modes. Hence, the RPs show more anharmonicity compared to CPs.
Additionally, the anharmonic coefficient derived from the FWHM of phonon modes, B (Eq.\ref{eq:5p4}) is also more for Raman phonons than for coherent phonons (bottom panel of Figure \ref{Fig:5p7}). The comparison of peak frequency and cubic anharmonic strengths for Raman and coherent phonon modes in various materials, such as pyrochlore titanates Dy$_2$Ti$_2$O$_7$ and Gd$_2$Ti$_2$O$_7$ \cite{Kamaraju2011}, as well as metal Bismuth \cite{Lu2018,Ishioka2006} and semiconductor CsPbCl$_3$ \cite{Nedelec2003,Nemec2005}, show similar trend that coherent phonon modes exhibit a weaker anharmonicity compared to corresponding Raman phonons. Further investigation is required to elucidate the underlying mechanisms driving this distinctive behavior of coherent phonons.


\section{Conclusions}
The optical pump optical probe spectroscopy of Ta$_2$NiSe$_5$ (TNSe) reveals the dynamics of carriers and coherent phonons. The carrier dynamics are primarily governed by the recombination of photoexcited carriers into excitons, which can be understood by the Rothwarf-Taylor model for the condensed phase of excitons in TNSe. The slow relaxation process represents the thermalization of phonons, which takes approximately 10 ps to cool down. The coherent phonon M3 shows deviations from anharmonic behavior for temperatures above $T_C$. Mode M2 exhibits a mean-field-like order parameter, supported with Raman measurements, while M3 shows asymmetric FANO lineshapes showing a change in temperature dependence across $T_C$. Coherent phonon modes show less anharmonicity compared to Raman phonon modes. The CWT analysis reveals that the formation time of quasiparticles ($t_{peak}$) is similar for all modes ($\sim$1 ps) except M3 ($\sim$3.5 ps). The temperature dependence of $t_{peak}$ describes the dressing of the M2 mode by excitons in the EI phase and illustrates the role of excitonic condensate for M3 mode across $T_C$. CWT analysis supports the temporal nature of the asymmetry of the M3 mode,  indicating that photoexcited carriers are the cause of this asymmetry. Based on the temperature-dependent behavior of asymmetry and the nature of the order parameter exhibited by the M2 mode, we assert that it is coupled to the excitonic phase of TNSe. In addition to the various signatures of the structural transition of TNSe observed in the carrier dynamics and the behavior of coherent phonons, this study reveals the time evolution of coherent phonons, providing new insights into the many-body effects.

\section*{Acknowledgments} 
AKS thanks the Department of Science and Technology for its financial support under the National Science Chair Professorship, and VA acknowledges CSIR for the research fellowship. VA acknowledges N. Kamaraju for an insightful discussion of the continuous wavelet transform procedure.



\clearpage
\newpage
\bibliographystyle{unsrt}
\bibliography{citations_Ta2NiSe5_P.bib}

\end{document}